# Negotiating Control and Accountability among Stakeholders in High-Risk AI Use Cases: Development and Evaluation of the Method ALIGN

Lena Schneider[1], Daniel Boos[2], & Gudela Grote[1]

[1] ETH Zurich, Department of Management, Technology, and Economics, Switzerland

[2] Swiss Federal Railways, Switzerland

Characteristics of Artificial Intelligence (AI) severely restrict human control. This creates new challenges for socio-technical systems design especially in safety-critical domains, where control over risks is essential along with the accountability for any harm caused. Drawing from human factors and AI governance literature, expert interviews and participant observation in a complex high-risk AI use case from the railways, we developed ALIGN, a scenario-based method that supports the governance of and stakeholder involvement in AI development and implementation in three ways: raising awareness for potential accountability-control misalignments in socio-technical systems and their negative consequences, offering a structured approach to assessing the alignment between accountability and control for stakeholders in a given use case, and supporting effective design decisions to mitigate AI risks. We describe the iterative development and evaluation of ALIGN and discuss how this method promotes AI governance by highlighting the relevance of control-accountability alignment throughout the AI life cycle.

**Keywords and Phrases: AI Governance, Human Factors, High-risk AI Use Cases, Accountability-Control Alignment**

## 1 INTRODUCTION

It is received wisdom that AI-based systems come with both great opportunities and challenges. Especially general purpose AI offers unprecedented capabilities to automate or augment ever more tasks but is also inherently opaque and only partially controllable due to the enormous complexity and self-learning capabilities of the models employed [46]. Moreover, the AI supply chain, involving the developers of certain applications along with companies that provide data, models, and computing power as well as a range of subservices such as data annotation and testing or fine-tuning of models, results in a multitude of interdependencies among a wide range of actors that are difficult to oversee and control, also called the "many hands problem"[35]. As a consequence, AI governance, that is the sets of rules by which all activities in the AI life cycle are directed and controlled, and all actors' decision-making authority and accountability are determined [47], is a key concern and has spurred a multitude of regulatory efforts, foremost the EU AI Act [45]. To date, regulations have focused on high-risk AI systems that have the potential to cause major harm to people's health, safety, or fundamental rights. How fine the line can be between preventing and causing harm has been shown in the case of Anthropic's powerful model Claude Mythos Preview that has raised serious cyber security concerns as it is capable of detecting and exploiting previously undetected vulnerabilities in operating systems and browsers and complete complex cyber operations with little to no human expertise required, enabling potential bad actors to execute cyber attacks [48]. Mythos demonstrates how AI capabilities advance faster than the ability for governance[49]. These fastly growing capabilities of AI models therefore show more than ever how urgently questions of accountability and control have to be addressed in the context of AI.
In light of these complexities inherent in AI development and use, regulators and companies struggle to define and install adequate measures for controlling AI-related work processes and to assign accountability accordingly. Mostly, end users of AI systems are considered accountable for assuring human oversight over system functioning and outcomes, as for instance in the EU AI Act (see for example [45,50]), likely creating once more the long-known "ironies of automation" [66] by holding humans accountable for managing risks in technologies that were designed to outperform them. However, shifting accountability to developers also has its challenges as they themselves do not anymore have full control over the systems they build [67], as shown by recent international headlines: Several employees of large technology companies have made public statements about AI being out of control and potentially killing all humans in the near future [52, 53,54]. In line with these discussions, the chief scientist of OpenAI, Jakub Pachocki, criticized that no AI company so far has developed appropriate alignment and monitoring capabilities to justify continuing to scale AI systems in a responsible manner [51].

This paper addresses this conundrum by proposing the method ALIGN which aims to support stakeholders in identifying and rectifying misalignments between control and accountability during AI development, implementation, and monitoring. ALIGN was developed and validated as part of a project in which an AI system for visual inspection of safety-critical components of cargo wagons is to be designed. However, as it is based on scenarios from multiple industries it is also applicable throughout different industries outside of this particular use case and of rail transport more generally, with a particular application focus on safety-critical domains. In the following we provide the conceptual principles on which ALIGN rests and subsequently describe the method itself and the empirical steps taken to develop and validate it.

## 2 BACKGROUND

### 2.1 Control and accountability as key components of good governance

Governance in the broadest terms refers to the rules in place for defining who is in control of what in any organization or institution and who is accountable for any actions taken (see for example [68]). In agency theory, a fundamental governance problem is discussed which concerns the alignment of interests between principals, such as firm owners, and agents, such as the firm's managers (see for example [69]). Principals assign accountability to agents in order to bind them to achieving outcomes that are in the interest of principals, or more generally some superordinate entity. Often not explicitly stated, the agents are assumed to have the necessary control to achieve those outcomes. Thus, control and accountability are aligned at the level of the agents, enabling and motivating them to do the right thing. The organizational literature and stakeholder governance research are full of examples where this alignment fails, due to power imbalances and the sheer complexity of the work processes to be governed [see for example 70, 73, 74]. In the realm of emerging technologies, Tesla's refusal to be even partially held to account for accidents caused by its autopilot [71] and the long-standing debate on the control of providers of digital labor platform over the presumed "freelancers" are cases in point [72]. Those examples also show how firms face negative consequences - at least in terms of reputational damage - if they are unable to achieve alignment, demonstrating its crucial role in governance.

## 2.2 Aligning control and accountability in AI development and use

By focusing on control and accountability and how it is (mis)aligned in the context of AI, key principles of good governance are brought into the framing of AI governance and the mitigation of risks through that governance [11]. Such an approach does not replace the classic components of risk management [55,60,61], that is risk identification, assessment, and mitigation. Rather, it provides a fundamental common language to discuss sources of risks and how they can be best managed. Moreover, the generic concepts of control and accountability help to address governance issues long before they may surface in the actual use of a technology and identify control and accountability requirements for technology developers to be able to prospectively resolve those issues. Thereby plausible futures, both desired and undesired ones, can be considered and acted upon before they materialize [62, 63].

For the purpose of the method ALIGN, we have built on definitions of control and accountability that cater for the specificities of technology development and use [3]. Control, defined as the capability to achieve desired outcomes includes three components transparency, predictability and influence [3, 10, 29]. Actors are in control when they have sufficient decision-making authority to influence situations in line with their goals and have adequate knowledge to effectively use their influence, which requires transparent ongoing processes and predictable future states and outcomes. Accountability refers to an actor's obligation to explain and justify their actions towards others [4, 28]. Three kinds of accountability can be distinguished: visibility, which refers to the demand to provide an intelligent account of one's own actions to others [39]; responsibility, referring to an actor's obligation to perform certain tasks in line with their formal work role and the specific allocation of functions among actors actions in line with ones' tasks or roles to achieve organizational goals [64]; liability, relating to requirements for an actor to answer to law, regulations, and contracts [65]. Liability becomes salient after failures (i.e., accidents) or misuse of a system (i.e., violation of data protection), while visibility and responsibility are relevant during regular work activities.

With increasing automation, human actors become supervisory controllers of systems [75], meaning that they monitor the functioning of systems and only intervene when problems arise, which in recent regulations has been termed human oversight. The less transparent and predictable the technical systems and the shorter the given timeframes for interventions are, this control becomes illusory. However, in their role as supervisory controllers, humans are still held accountable for system functioning and outcomes. This control-accountability gap widens in the context of AI, especially due the complexity and self-learning capabilities of AI systems, which drastically reduce transparency and predictability not only for AI users, but also for AI developers. Giving a Large Language Model the same prompt twice is very likely to generate two different answers, with the user having to check the veracity of either, for which deep knowledge of the respective domain is required.

Additionally, in complex AI use cases with many different stakeholders, the boundaries of social-technical systems widen to what Slota et al. [35] have termed "socio-technical landscapes" where the distribution of control and accountability is increasingly difficult to determine and to trace over time through the different phases of an AI life cycle. Importantly, the task for good AI governance reaches beyond the alignment of control and accountability for end users by encompassing all relevant stakeholders, from data annotators to regulators [32, 58]. This raises questions as varied as "Do data annotators working through digital labor platforms have the working conditions and the required knowledge to provide high-quality data for model training?" or "How is accountability for AI safety distributed between regulators, system developers, and system operators, that is, are prescriptive or process-oriented regulatory regimes [76] followed?"

Failing to address issues of control-accountability alignment proactively can have severe consequences as actors aim to push accountability, particularly liability after system failures, on to other actors, and end users as the least powerful stakeholders are likely to be blamed by more powerful actors (i.e., executive management) for consequences of AI they had no influence over. These dynamics can severely hamper the safe functioning of systems and the necessary learning and adaptation when systems fail. It is therefore indispensable for risk mitigation to effectively address potential control-accountability gaps as early as possible and continue to do so through the AI life cycle [11]. Accountability should be designed into AI systems and their surrounding governance structures from the start [59] and the necessary organizational capabilities built up well in advance before any AI system is integrated in safety-critical contexts [60]. Stakeholder participation is crucial to achieve these objectives, to which we turn next.

### 2.3 Stakeholder involvement in AI development

In order to address the distribution of control and accountability proactively and effectively, all relevant stakeholders need to be involved and guided through complex and potentially highly political negotiations. Participation of end users during later stages of technology design is common practice, however with AI, everyone agrees that participation has to start earlier and involve more stakeholders [7, 9, 23, 24, 34, 43, 45]. Stakeholders typically include various internal and external actors such as developers, data scientists, domain experts, managers, operators, and regulators [27]. Each group holds distinct forms of knowledge - technical, procedural, and contextual - that must be integrated to achieve systems that are both effective and acceptable in practice [16]. Involving a broad range of stakeholders can support balancing decision power and lead to better understanding of the work system that the AI will be introduced in, as well as a more accurate prediction of risks [58].

To date, concrete and actionable guidance on how to achieve such broad participation is lacking [16, 17]. For instance, end users as experts on the actual work processes are usually not included in decision making on the functionality of systems, and developers may fail to effectively communicate requirements for training data quality or quantity to organizations. Similarly, secondary stakeholders are rarely involved in design processes at all [77]. To allow integrative negotiations on control-accountability alignment in a broad network of stakeholders, where actors are willing to work towards equitable solutions rather than follow winner-takes-it-all tactics, participatory approaches have to foster common visions of a future AI-enabled work system based on extensive information sharing and perspective taking [11]. Existing methods for participatory design, defined as "activities where users, designers and researchers collaborate towards shared goals" [2], aim to draw from the experiences, preferences, capabilities and limitations of future end users of technologies and other stakeholders in order to inform design decisions and foster usability [31, 36]. Function allocation between technology and human actors may also be a topic of co-design with stakeholders, however methods for addressing the underlying fundamental questions of who is in control and who is accountable, also in view of potential negative consequences of AI systems, are rare [3, 7, 14].

In search of an appropriate participatory design method for control-accountability negotiations, workshops are assumed to be a useful approach as different perspectives by a range of stakeholders can be articulated and new shared understandings developed [31]. Workshops create structured opportunities for stakeholders to jointly reflect on, negotiate, and re-assign control and accountabilities. Especially in early stages of technology development, such workshops will not focus on co-design but provide spaces in which control and accountability can be explicitly discussed and reconfigured and solutions to potentially highly contested governance problems co-created. Unlike traditional requirements elicitation techniques, workshops enable multi-directional learning between developers, users, and other stakeholders, thereby fostering shared mental models of the work system, the AI embedded in it, and the challenges for control and accountability arising during AI development and use [5]. Lastly, workshops reflect themselves an appropriate alignment of control and accountability by giving stakeholders an active role in shaping AI systems and increasing participants' influence over decisions that affect their future accountability.

In the following, we present the work undertaken to develop and evaluate the method ALIGN which aims to support negotiations about control-accountability alignment during AI development and use for high-risk systems.

## 3 DEVELOPMENT OF THE METHOD ALIGN

### 3.1 Context

ALIGN was developed alongside the first authors' research conducted within a larger research project approved by the university ethics board and partially funded by a railway organization. The first author studied a complex high-risk AI railway use case within this organization, where a new AI-based system will support experts operating from a control center in detecting damages on cargo trains and initiating appropriate measures in close collaboration with other human actors (i.e., dispatchers, maintenance coordinators, maintenance technicians). Images of cargo trains taken during normal operation will be processed by the AI-based system. In case of any identified damages, the system will alert the control center expert who has to make decisions about the prevalence of a damage and prioritize measures. This system is assumed to benefit railway operations in a number of ways: improved worker safety by moving technical inspectors from the tracks to the control center; reduced maintenance costs due to early damage detection; less vulnerability to shortages of qualified workers; improved efficiency of maintenance processes by optimizing maintenance schedules; increased punctuality of trains.

In addition to the beforementioned stakeholders involved in the use of this system, there are multiple actors involved in the development, governance, licensing and implementation, including project leaders, executive managers, safety experts, human factors experts, developers, infrastructure experts, future end users, and regulators. These actors differ widely in their professional backgrounds, work tasks, organizational association and the involvement with the use case, and accordingly also differ widely in their capabilities and expectations regarding the assignment of control and accountability to themselves and to others. Common practices in project management do not adequately cover new demands for collaboration among such a wide range of stakeholders whose roles may also change over the course of the AI life cycle. For instance, end user may have to be involved very early on as annotators of training data, rather than only during final stages of interface design and system implementation. With the increasing complexity of AI use cases, practitioners, organizations, and government agencies require tools that allow them to involve more stakeholders earlier and in more fundamental design decisions. ALIGN can therefore be useful well beyond the described use case.

### 3.2 Purpose of ALIGN

ALIGN aims to support stakeholders in complex high-risk AI use cases during negotiations about the assignment and alignment of control and accountability, building on concepts from AI governance, work design, and stakeholder participation. It thereby differs from existing participatory methods that either focus on more narrow questions of user-centered design (e.g., HAX toolkit by Microsoft [57]), such as explainability and interpretability of AI systems and their output, or reach out to large stakeholder groups to capture general AI-related expectations without reference to a specific ongoing AI project (e.g., [58]. A review by Delgado et al. [7] also found that participatory approaches often involve stakeholders only during short periods after key design decisions are already made.

ALIGN was developed in close collaboration with practitioners to ensure its practicability in terms of fit to existing organizational practices, such as agile product development and conformity with technical process standards, and to existing tools for digital innovation. Also, the method was designed to address the needs of multiple and very heterogeneous stakeholders operating in complex work systems by involving them directly or their representatives. The main focus of ALIGN is to support stakeholders in control-accountability negotiations at various stages of AI development and implementation, however, it can also serve as documentation of the design process to meet obligations related to internal governance and regulatory compliance.

### 3.3 Structure and Materials

Core to ALIGN is a moderated workshop conducted in person with groups of three to eight participants (stakeholders of the respective AI use case and/or their representatives). To lower the barrier to use the method, we designed the workshop to take significantly less than half a workday, about 2 to 2.5 hours. ALIGN can also be used to discuss questions of control and accountability alignment with single stakeholder groups, which can deepen their understanding of the issues involved and prepare them for negotiations within the larger network of stakeholders.

The workshops can be facilitated by project leaders, project members, or also third parties, such as external moderators. Each option has its advantages and disadvantages in terms of knowledge and the individual and collective stakes involved so that the choice should be made carefully and possibly also revisited for further ALIGN workshops over the duration of the project.

The ALIGN workshop is divided into three parts: a brief theoretical introduction, a discussion of three to five scenarios from diverse industries with the help of a scenario card set which each include different kinds of misalignments between control and accountability, and lastly a discussion of the control-accountability fit for different stakeholders within the participants' own use case. The fit assessment is supported by an AI stakeholder canvas as a visual facilitation tool and is based on the current state of design decisions already taken or about to be taken in the project. Any detection of misfits is then used to develop design ideas to improve alignment. Table 1 shows the structure of the workshop including the content, materials, goals, tasks for the facilitator, and duration for each workshop part.

Table 1. Overview of ALIGN workshops

| | **Part 1** | **Part 2** | **Part 3** |
|---|---|---|---|
| Content | Theoretical introduction | Scenario-based discussion of control-accountability misalignment in diverse industries | Discussion of control- accountability alignment in use case |
| Materials | Slide deck | Scenario cards, flip chart/white board, sticky notes | AI stakeholder canvas |
| Goal | Provide a brief overview of the topic of the workshop and define concepts of accountability and control | Initiate discussions around different types of control-accountability misalignments in concrete situations and collect initial ideas to address them | Create an overview of control and accountability requirements for all stakeholders in use case, create awareness for and identify potential misalignments, derive initial design ideas to ensure control-accountability alignment, create an overview of requirements for control-accountability alignment as to be followed up during regular project work |
| Tasks of the facilitator | Introducing and defining concepts, explaining and illustrating their relevance | Choosing scenarios for the discussion, moderating discussion with the help of guiding questions and follow up questions, collecting and organizing design ideas from participants, time keeping | Introducing the visual facilitation tool and instructing participants how to use it, encouraging participants to identify solutions for addressing misalignments, time keeping |
| Duration | Approx. 15 minutes | Approx. 45-60 minutes | Approx. 60 minutes |

In the following, the materials of ALIGN are described in detail.

### *3.3.1 Slide deck*

A slide deck supporting a brief (approximately 15 minutes) theoretical introduction to the topic of accountability-control alignment in high-risk AI use cases was created. It is meant to support workshop facilitators in providing an overview of the workshop structure and content, underline the relevance of the topic for responsible AI development as well as introduce and define the control and accountability concepts.

### *3.3.2 Scenario card set*

A set of 19 scenario cards is provided for the second part of the workshop. Each card contains a short description of a fictitious scenario where a misalignment between control and accountability during AI development or use in a safety-critical domain led to negative consequences (i.e., accidents, delays, cost increases). The scenarios described on the cards vary with regard to the addressed safety-critical domain, the involved stakeholders, and the affected control and accountability aspects. While real-life discussions between stakeholders often focus on questions of liability, which becomes relevant mostly in the context of system failures, we wanted to ensure that participants also discuss the two other kinds of accountability, i.e. visibility and responsibility, as they are more relevant during regular operations. The safety-critical domains covered in the scenarios are railways, medicine/healthcare, air traffic control, and maritime transport. On the front side, the scenario cards contain short descriptions of the situations. On the back side, the cards contain information on the safety-critical domain addressed in the scenario, the involved stakeholders, and the relevant control and accountability aspects. The different domains are represented by different colors and icons. The information on the back of the cards is provided to support facilitators in choosing the three to five most suitable scenario cards for their workshop. Figure 1 shows an example of a front and back side of a scenario card. To facilitate discussions about control and accountability, workshop participants are first confronted with scenarios from domains other than their own, as prior research in participatory and scenario-based design indicates that neutral, decontextualized scenarios can enable participants to address sensitive issues without the constraints of existing hierarchies or interpersonal dynamics [6, 41, 56]. Therefore, attention was paid to describing situations in a way that is understandable also to non-domain experts to make all scenario cards usable across domains. Cards are meant to be chosen and read out loud by the workshop facilitator. Afterwards, workshop participants are instructed to discuss how the described situation from the scenario card could have been avoided through design decisions that would have better supported control-accountability alignment. Ideas are then written on sticky notes and collected on a whiteboard or flip chart. The scenarios are the core of the workshop, as they provide concrete situations that enable stakeholders to discuss various otherwise rather abstract governance issues. Due to the newness of AI, most stakeholders in safety-critical domains have limited or no previous experience in how to best develop, implement, and regulate these new technologies. Risks are often vague and control and accountability are abstract topics that are rarely explicitly addressed. By providing concrete examples of failures associated with ineffective decision-making that led to control-accountability misalignment for one or multiple stakeholders, ALIGN provides a shared, tangible reference point for discussions on AI governance and concrete design choices. Figure 1 shows examples of the front and back side of a scenario card.

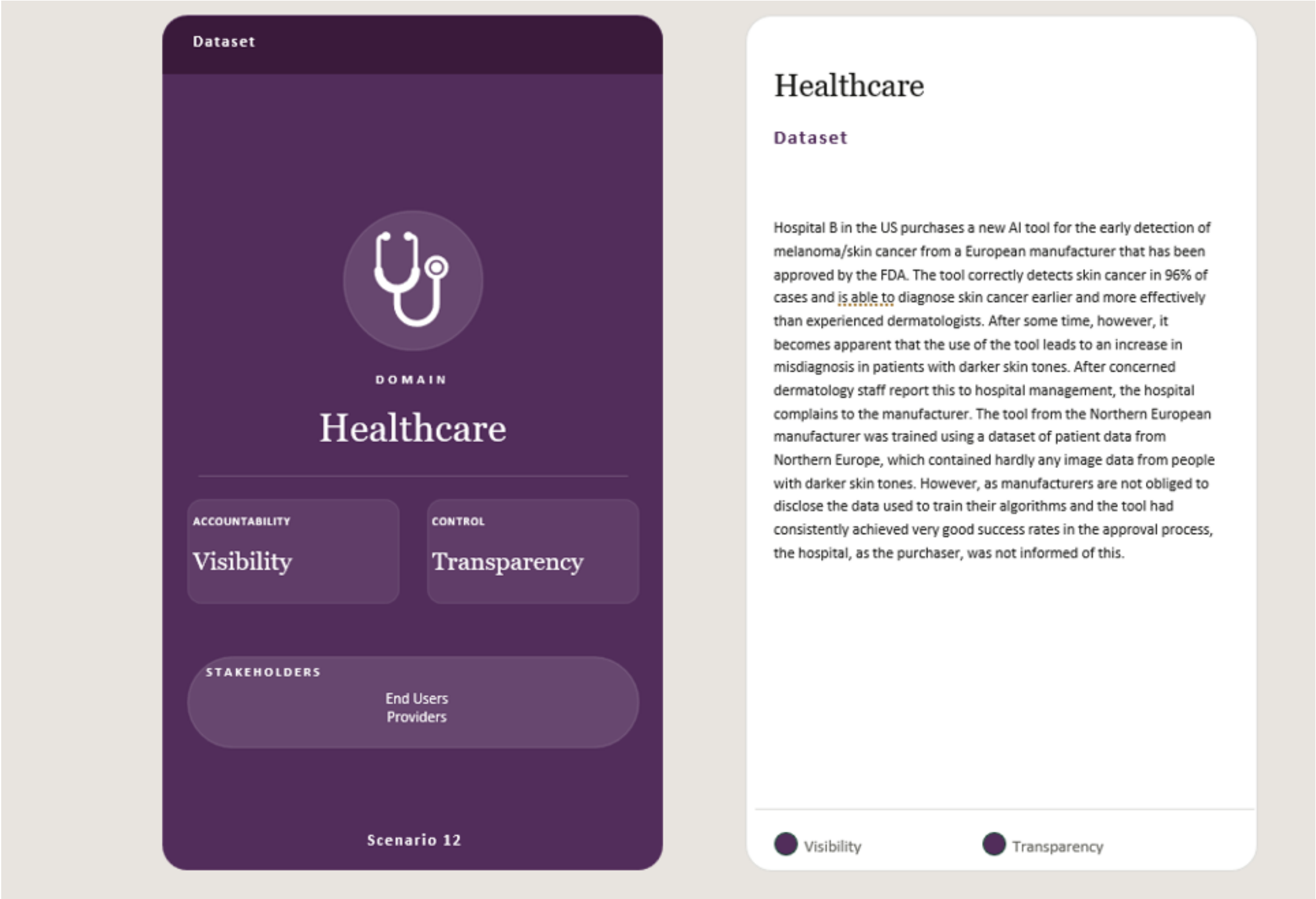

Figure 1: Example of the front and back side of a scenario card.

### *3.3.3 AI stakeholder canvas*

To facilitate the transfer of the scenario discussions to the workshop participants' own AI use cases, a visual facilitation tool in form of an AI stakeholder canvas was created to identify all relevant stakeholders in a particular AI use case and assess the potential (mis-)fit of control and accountability for each of them. The canvas was created inspired by existing visual stakeholder management tools. It allows stakeholders to explicitly list control and accountability for a role and assess the fit between them in an efficient way using a clear dichotomous categorization (low fit, indicating the need for further negotiations and adjustments and high fit, indicating no immediate need for improvements). Thereby, an overview of relevant stakeholders in the respective AI project is provided, listing their control and accountability capabilities and requirements, which creates a base for discussion amongst stakeholders, helping to address the “many hands problem” [35] in AI projects and fostering stakeholders' knowledge about each other. The canvas is shown below in Figure 2.

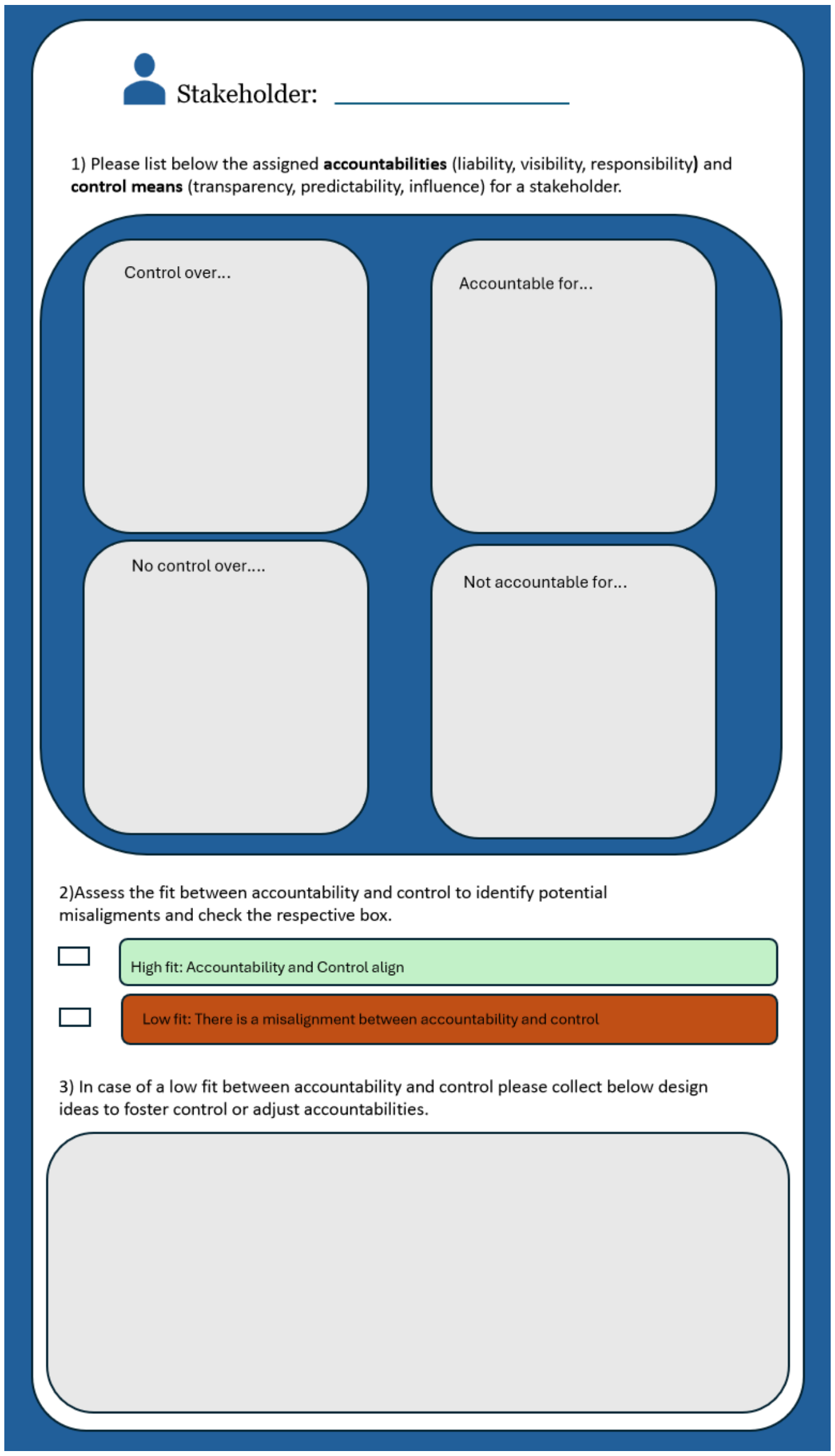

Figure 2. The AI stakeholder canvas to be used during workshops for accountability-control fit assessment.

Participants are instructed by the facilitator to fill out the four boxes on canvas for a respective stakeholder role by listing both what stakeholders are and are not in control of and are and are not accountable for in the respective box. After completing this step, participants then assess the fit between existing control means and accountabilities for each stakeholder. High fit indicates that no additional design measures need to be taken to improve control-accountability alignment, whereas a low fit indicates either too little control to fulfil the assigned accountabilities or too much control without matching accountabilities, both warranting further design decisions to optimize alignment. It should be noted that both high control/high accountability and low or no control/low or no accountability constitute alignment. Participants then present their assessment for each role to each other, which can reveal different perspectives on assigned accountabilities and control needs and spark discussions within the group. Whenever a low fit is identified, initial design ideas are collected within the group of participants to increase alignment and documented on the canvas. Participants can both explore the value of design ideas developed during the previous discussions of scenario cards or come up with additional ideas. Figure 3 shows a human factors expert using the canvas during the third part of the workshop. The results of the workshop, especially the overview created of the control means and accountabilities for each stakeholder as well as the design ideas, serve as a basis for more in-depth discussions and decision making during future project meetings and support the project team in reallocating control and accountability between stakeholders when needed, while also taking into account interdependencies between stakeholders. Clearly, such design decisions need to be communicated to stakeholders outside of the project team, and further workshops based on ALIGN are likely needed over the course of development and implementation of the AI system. ALIGN can also help to resolve control-accountability misalignment once it becomes manifest during the use of the new system.

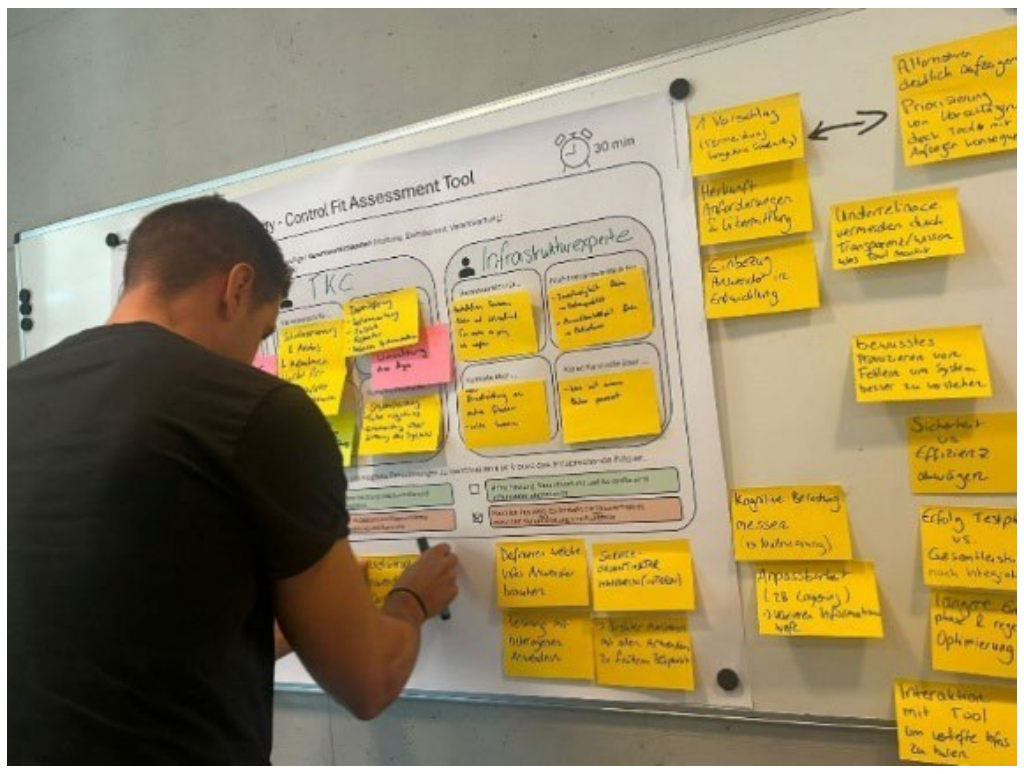

Figure 3: A human factors expert using the AI stakeholder canvas during the third part of a workshop. Photo taken by the first author.

### 3.4 Development and validation of the ALIGN scenarios

The scenarios are at the core of ALIGN. A scenario-based approach was chosen because scenarios not only enable discussions about and assessment of and changes to the design of technologies before they even exist, but also allow participants without prior knowledge or expertise to engage in discussion around design (see for example [56]), making this approach particularly suited for the purpose of ALIGN. The ALIGN scenarios were created in two ways, depending on the domain: For the railway scenario cards, situations were derived from interviews with stakeholders within a real-life AI use case and observation of collaboration between these stakeholders in project meetings over the duration of two years from 2023 to 2025. For the other safety-critical domains, scenarios were derived based on literature search and interviews with different domain experts (i.e., physicians, air traffic controllers, human factors experts from different high-risk organizations), focusing on current challenges within the specific safety-critical domain regarding AI governance and (potential) harm caused by AI. We then identified those situations where the described risks or negative outcomes could be traced back to a control-accountability misalignment. The situations were reviewed by domain experts in terms of their realism, relevance, and wording. The reviewers provided written feedback including concrete suggestions for editing. After final editing of the scenarios, two independent individuals assigned the appropriate control and accountability aspects the scenarios addressed. One scenario had to be rephrased to clarify the relevant control and accountability aspects. While the scenario card set currently consists of 20 cards, it is designed with the idea in mind that more AI scenarios can be added over time by organizations that conduct the workshops, as their experience with AI use cases grows and relevant examples arise. The process for the development and validation of the scenarios is shown below in Figure 4.

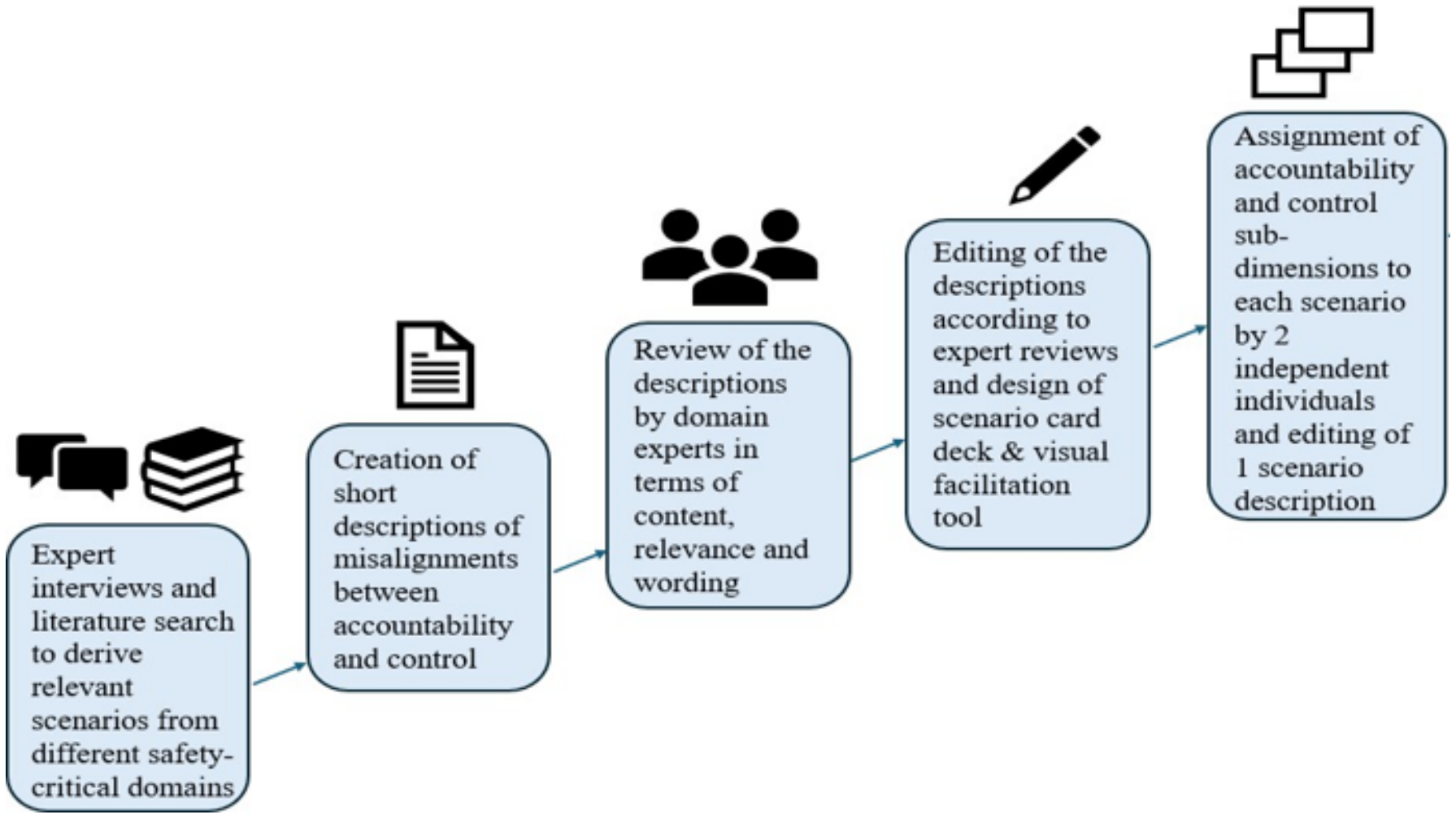

Figure 4: Overview of the development and validation process for the ALIGN scenarios.

## 4 EVALUATION OF ALIGN

We followed a mixed-method approach for our evaluation of ALIGN, combining quantitative and qualitative methods. The final version of the scenario card set, together with the introductory slide deck and the canvas were applied in five workshops that were facilitated by the first author. Participants for the workshops were recruited through the research project within the railway company as well as personal contacts of the researchers. All participants provided informed consent and participation in the workshops was voluntary and unpaid, with no material or financial incentives or rewards being offered. The research was approved in advance by the university's ethics board (approval number withheld for anonymity reasons at this point in time, can be provided upon acceptance).

In the first workshop, we collected qualitative feedback focusing on workshop structure, feasibility and clarity of instruction, and understandability of materials. It was conducted with three work and organizational psychology researchers as participants. As we wanted to ensure the workshop can be run across safety-critical domains with participants with different backgrounds, including regulators or executive level managers who either might have less detailed knowledge of the work system the AI-based system will be implemented in, or little technical knowledge about AI, an evaluation with non-experts in terms of domain and AI knowledge provided an opportunity to test whether the materials, particularly the content of the scenario cards, were understandable. The workshop was run on site at the research university the first and third author are employed at. For the last part of the workshop, participants were provided with role cards describing three stakeholder roles based on the real-life visual inspection use case that inspired this method (a technical inspector as an end user; a developer, and an infrastructure expert) and asked to fill out the canvas for these three stakeholders. One stakeholder role description represented a high fit between control and accountability, while the other two represented misalignment. Design ideas were generated to improve alignment for these two latter roles. Post-workshop feedback was obtained in a focus group discussion, and the workshop materials and instructions were edited according to the feedback. The updated version of ALIGN was used in the subsequent workshops.

The second workshop was conducted with four human factors and UX experts from the railways. These participants not only possessed detailed knowledge of socio-technical systems, but also of the railway domain and were closely involved in projects in which new technologies, including AI, were introduced. After the workshop, we collected individual feedback using an online survey with MentiMeter and discussed participants' perceptions of the workshop in a focus group. Participants were asked to provide feedback with regard to the structure, content, materials, and relevance of the workshop as well as the level 1 (satisfaction) and level 2 (learning) of Kirkpatrick's model [18] of training evaluation, which can be used for evaluating other types of interventions as well (i.e.,[30]). As this workshop was conducted outside of one specific real-life AI project, it was not feasible to include the other two levels of evaluation (behavior and results). This approach is common practice since most intervention studies focus on evaluating level 1 and 2, given the complexity of measuring actual behavior change and the results of that behavior change (see for example [25, 38]). For the online survey on participants' satisfaction and learning, all items were rated on 5-point Likert scales ranging from 1 (*"not at all"*) to 5 (*"completely"*). The items were loosely based on [25]. Additionally, we asked about the match between the workshop and existing tools or methods used within the railway organization and the applicability within the railway organization. Lastly, participants were given the opportunity to add additional comments with regard to potential improvements in open text fields.

The third workshop was conducted with five AI ethics experts from the same railway organization. The content and structure of the evaluation were the same as for the second workshop. As these experts are familiar with the concept of accountability, an evaluation with this group of participants allowed additional insights into the relevance of scenarios and their appropriateness for initiating discussions and deriving design ideas.

The fourth workshop was run with four different stakeholders of the visual inspection use case from the railways that inspired the method development, and evaluated the same way as workshop 2 and 3.

Lastly, to further evaluate the applicability of ALIGN in safety-critical domains other than the railways, it was used in a fifth workshop with four employees of a regulatory agency for nuclear energy, where only qualitative feedback was collected on the relevance and appropriateness for this domain as well as perceived effectiveness of the method for addressing control-accountability alignment during a focus group after the workshop.

### 4.1 Quantitative evaluation results

Across the workshop rounds, participants rated the materials, content, and structure of ALIGN favorably, with most items averaging above four on the 5-point scale. The highest possible rating was given by all participants for the workshop content being of interest to them. Ratings were also high across groups for ALIGN's ability to initiate and foster discussion and reflection on control-accountability alignment. The only item with a mean below the scale midpoint was a negatively formulated item stating the introduction was too long, which participants disagreed with. Table 2 below presents the quantitative evaluation results combined across all workshops and for each workshop individually.

Table 2: Participant ratings of workshop materials, structure and content, as well as instructions, match with existing company tools, and self-perceived knowledge gain, combined across all workshop rounds and broken down by individual workshop. All ratings are on a 5-point Likert scale from 1("not at all") to 5 ("completely").

| **Evaluated Aspect** | **Combined across Workshops (*M, SD*)** | **Workshop 2 (*M, SD*)** | **Workshop 3 (*M, SD*)** | **Workshop 4 (*M, SD*)** |
|---|---|---|---|---|
| **Number of participants** | **12** | 4 | 5 | 4 |
| **Workshop Materials** | | | | |
| Usefulness of materials overall | 4.33 (*SD*=0.49) | 4.25 (*SD*=0.50) | 4.40 (*SD*=0.55) | 4.33 (*SD*=0.58) |
| **Scenario Cards** | | | | |
| Scenario descriptions were clear | 4.67 (*SD*=0.49) | 4.50 (*SD*= 0.58) | 4.60 (*SD*=0.55) | 5.00 (*SD*=0.00) |
| Scenarios were relevant | 4.83 (*SD*=0.39) | 4.75 (*SD*= 0.50) | 5.00 (*SD*=0.00) | 4.67 (*SD*=0.58) |
| Use of cards was helpful | 4.33 (*SD*=0.78) | 5.00 (*SD*=0.00) | 4.20 (*SD*=0.84) | 3.67 (*SD*=0.58) |
| Number of cards used was appropriate | 4.08 (*SD*=0.79) | 4.00 (*SD*= 0.82) | 5.00 (*SD*=1.00) | 4.33 (*SD*=0.58) |
| **AI Stakeholder Canvas** | | | | |
| Was clearly structured | 4.25 (*SD*=1.14) | 3.50 (*SD*=1.73) | 4.80 (*SD*=0.45) | 4.33 (*SD*=0.58) |
| Was helpful for working towards workshop goals | 3.92 (*SD*=1.17) | 3.75 (*SD*=1.89) | 4.00 (*SD*=1.00) | 4.00 (*SD*=0.0) |
| Was important for reflection on control-accountability alignment | 4.58 (*SD*=0.52) | 4.75 (*SD*=0.50) | 4.40 (*SD*=0.55) | 4.67 (*SD*=0.58) |
| **Theoretical Introduction** | | | | |
| Contained an appropriate level of detail | 4.75 (*SD*=0.62) | 4.75 (*SD*=0.50) | 5.00 (*SD*=0.00) | 4.33 (*SD*=1.16) |
| Was relevant | 4.75 (*SD*=0.45) | 4.75 (*SD*=0.50) | 5.00 (*SD*=0.00) | 4.33 (*SD*=0.58) |
| Was clear | 4.25 (*SD*=0.87) | 4.00 (*SD*=0.82) | 4.40 (*SD*=0.89) | 4.33 (*SD*=1.16) |

| **Evaluated Aspect** | **Combined across Workshops (*M, SD*)** | **Workshop 2 (*M, SD*)** | **Workshop 3 (*M, SD*)** | **Workshop 4 (*M, SD*)** |
|---|---|---|---|---|
| Was too long | 1.92 (*SD*=1.56) | 2.5 (*SD*=1.92) | 2.20 (*SD*=1.79) | 1.00 (*SD*=0.00) |
| **Workshop Instructions** | | | | |
| Were clear | 4.17 (*SD*=0.72) | 4.00 (*SD*=0.82) | 4.20 (*SD*=0.84) | 4.67 (*SD*=0.58) |
| Were helpful | 4.50 (*SD*=0.52) | 4.75 (*SD*=0.50) | 4.60 (*SD*=0.55) | 4.00 (*SD*=0.00) |
| Contained an appropriate level of detail | 4.58 (*SD*=0.52) | 4.75 (*SD*=0.50) | 4.60 (*SD*=0.55) | 4.67 (*SD*=0.58) |
| **Workshop Overall** | | | | |
| Inspired discussions on control-accountability alignment | 4.83 (*SD*=0.39) | 4.75 (*SD*=0.50) | 4.80 (*SD*=0.45) | 5.00 (*SD*=0.00) |
| Supported reflections on the topic | 4.75 (*SD*=0.45) | 4.75 (*SD*=0.50) | 4.80 (*SD*=0.45) | 4.67 (*SD*=0.58) |
| Was helpful | 4.50 (*SD*=0.67) | 4.25 (*SD*=0.96) | 4.80 (*SD*=0.45) | 4.33 (*SD*=0.58) |
| Was interesting | 5.00 (*SD*=0.00) | 5.00 (*SD*=0.00) | 5.00 (*SD*=0.00) | 5.00 (*SD*=0.00) |
| **Applicability of ALIGN within Railway Company** | | | | |
| Match with existing tools within the railway company | 4.30 (*SD*=0.95) | 4.00 (*SD*=1.16) | 4.75 (*SD*=0.50) | 4.00 (*SD*=1.00) |
| Relevance for railway company | 4.58 (*SD*=0.52) | 4.25 (*SD*=0.50) | 4.60 (*SD*=0.55) | 5.00 (*SD*=0.00) |
| Feasibility of repeating workshop at the railway company in the future | 3.82 (*SD*=1.08) | 3.25 (*SD*=1.50) | 4.50 (*SD*=0.58) | 3.67 (*SD*=0.58) |
| **Self-Perceived Knowledge Gain** | | | | |
| Something new learned during the workshop | 3.75 (*SD*=0.71) | - | 4.00 (*SD*=0.71) | 3.33 (*SD*=0.58) |
| Learned more about the concepts of control and accountability | 4.00 (*SD*=0.54) | - | 4.20 (*SD*=0.45) | 3.67 (*SD*=0.58) |

| Evaluated Aspect | Combined across Workshops (*M*, *SD*) | Workshop 2 (*M*, *SD*) | Workshop 3 (*M*, *SD*) | Workshop 4 (*M*, *SD*) |
|---|---|---|---|---|
| New knowledge about control–accountability alignment gained | 4.13 (*SD*=0.64) | - | 4.40 (*SD*= 0.55) | 3.67 (*SD*=0.58) |

[a] *M= Mean; SD= Standard Deviation. No quantitative data was collected in workshop 1 and 5 respectively due to their focus on qualitative feedback on feasibility, clarity and relevance outside of the railway sector.*

As can be seen in the table above, the AI ethics experts participating in the third workshop were more confident in regards to the feasibility of running ALIGN workshops within the railway company in the future than both the human factors experts in workshop 2 and the stakeholders of the real-life AI use case in workshop 4. This matches the discussions during the focus groups, where the AI ethics experts identified multiple ways to foster the use of ALIGN and incorporate it in the company's AI governance (see chapter 4.2).
The human factors experts from the second workshop saw the use of cards during the workshop as more helpful than the stakeholders from the other workshops. The perceived knowledge gain also varied: stakeholders from the 4$^{th}$ workshop, who had been involved in their own AI use case for several years and were previously confronted with accountability-control issues that sparked the development of ALIGN in the first place, gave lower ratings for having acquired new knowledge than the AI ethics experts, while at the same time still providing the highest possible rating for the relevance of ALIGN for the railway company overall.

### 4.2 Qualitative evaluation results

Comments by workshop participants, both written and orally communicated, were overall very positive across all workshops. Feedback was particularly positive in regard to the relevance of the topic and to using the workshop as *"a great way to raise awareness"* (participant of the second workshop) of issues resulting from control-accountability misalignments and supporting AI governance: "*This [ALIGN] is very well thought out; for me, it would be relevant to our AI governance or has a lot of overlap with it"* (participant of the third workshop).
In regards to the structure and duration of the workshops themselves, all were successfully conducted within the desired time frame, and the facilitator was able to follow the desired agenda for all workshop parts and instruct participants as intended.
The part of the workshop that received the most positive feedback during focus groups were the scenarios: Participants of all workshops underlined the relevance and helpfulness of the scenarios, with several participants mentioning they would like to discuss more scenarios in the future, for example a participant of workshop 2 *("can we talk about some more of these [scenarios]sometime?")*, and positively speaking about the discussions they initiated, as done for example by a participant of the fifth workshop: "*I particularly liked the discussion that started within the group because of the scenarios"*. This is in line with the observation by the facilitator that participants were indeed able to engage in lively discussions about each scenario and create at least six design ideas for each scenario in every workshop without needing additional guidance or support, often creating many more ideas.

The use of scenarios from domains other than their own was positively evaluated as helpful and thought-provoking by participants. As *"accountability is a very emotional topic"* (Participant of workshop 2), discussing scenarios from other domains avoids potential conflict between workshop participants based on their role within their own use case. Yet, participants often transferred their design ideas to their own professional experience from their own use case ("*That would be the same for us"* - participant of workshop 5 discussing a scenario where unknown information about the training data set restricted the control of the regulators; *"we have to pay attention to that in our project as well"* participant in workshop 4). During the workshop with the real-life AI use case stakeholders, participants frequently referred back to the scenarios when presenting their assessments for each of their own stakeholder roles and discussing design ideas. The discussions following the accountability-control fit assessment for stakeholder roles during that workshop also indicated that, even though the stakeholders had collaborated within their use case over several years, there was still no shared understanding of how control and accountability had been assigned to stakeholders so far and as a consequence a wide range of perspectives on (mis)alignments existed ("*he [the end user] would be held accountable for [errors]"* (one participant of workshop four)) – *"no, he would not be held accountable, there's nothing he can do there, we said the company is accountable*" (another participant of workshop four)). At the end of workshop four, the participants discussed the next steps for further concretizing the ideas derived during the workshop during their regular project meetings, indicating that ALIGN had in fact raised awareness for the importance of negotiating control-accountability alignment and with that the motivation to invest effort into improving control-accountability alignment for the AI system they are working on.

The participants in the workshop at the nuclear regulatory agency also engaged in vivid discussions around the scenarios from all domains represented in the ALIGN card set, showing that a transfer to other safety-critical domains is feasible, and explicitly mentioning *"this is really, really helpful because we didn't even know where problems could arise with AI and what to pay attention to"* (participant of workshop 5). Furthermore, they discussed ideas on defining control and accountability for their own stakeholder role as well as for that of nuclear power plants. Participants not only expressed very high satisfaction with the workshop but also requested a second workshop to discuss more scenarios and invite additional employees of their institution. Subsequently, the workshop also informed the development of an internal white paper and provided the foundation for the formulation of a research proposal alongside a grant to further investigate AI governance in the nuclear energy sector from a socio-technical perspective. This project will allow the nuclear regulatory agency to build the capacity to proactively address control-accountability alignment long before AI systems are actually implemented in nuclear power plants.

The question of how AI project teams can be motivated to use ALIGN with the respective stakeholders was raised and discussed by the participants of the human factors expert group in workshop 2 and the members of the AI ethics group in the railway organization in workshop 3. Both groups stated that from their perspective, the current awareness of potential issues arising from control-accountability misalignment in AI development teams within their organization was still considered rather low as little previous experience with AI projects existed, while the relevance of addressing accountability and control was particularly high. Accordingly, the participants made several suggestions: Besides company-internal AI ethics campaigns and initiatives by the training and development teams related to fostering employees AI literacy and skills for future work tasks involving AI that could incorporate ALIGN, one promising approach might be to highlight ALIGN as a means to supporting documentation for regulatory processes, which is at the core of the efforts of project managers. Additionally, suggestions to *"provide the tool for self-service"* (participant of workshop 3) or *"to incorporate it in AI governance"*(participant of workshop 3) were made.

## 5 DISCUSSION

Our study aimed to address the challenge in the design of high-risk AI systems on how to align accountability and control among heterogeneous stakeholders in complex settings. By introducing and evaluating the participatory method ALIGN developed based on a real-world high-risk AI use case with the support of experts from several safety-critical domains, we address the concerns that existing guidance for responsible AI practice is too abstract (see Kallina et al., 2025). We show how with the help of ALIGN control-accountability misalignments as a key concern in AI governance can be identified and translated into design decision-making. While the combination of work and technology design remains challenging [15], we demonstrate how a simple method can be used to initiate and foster discussions about highly contested design decisions for AI systems in safety-critical domains.

The scenario cards developed for ALIGN provided concrete reference points for participants to ground discussions in realistic examples of control-accountability misalignment, while the visual facilitation tool supported identification of any potential misalignments in the stakeholders' own work context and spurred ideation on design solutions for better alignment. The scenario cards thus functioned as boundary objects [37] that can enable collaboration across highly diverse stakeholder groups [21, 26]. In complex AI projects, many diverse actors are involved, coming from different organizations (i.e., providers or regulators) and knowing very little about each other's work tasks and how those might be affected by AI. Gaining a deeper understanding of each other forms the basis for successful collaboration and in particular negotiations about sensitive issues involved in allocating control and accountability for the functioning and outcomes of AI systems.

With ALIGN, we provide an easy-to-use method to support the effective involvement of stakeholders in the development of AI systems. Traditional participatory design in HCI has emphasized user involvement to ensure usability, empowerment, and organizational fit [2, 33]. Extending this tradition, our method positions all stakeholders, not just end users, as participants in the design process and focuses on decisions related to both work and technology design. Additionally, the method can support the regulation of AI by involving regulators into control-accountability negotiations as a core element of AI governance within the limits of remaining independent oversight bodies and by documenting negotiated choices, thereby supporting compliance with emerging governance demands.

Successful control-accountability negotiations amongst stakeholders should also directly inform design decisions in high-risk AI projects: Once it has been decided how control and accountability are to be distributed amongst stakeholders, measures can be taken to ensure control of the respective stakeholders, both through technology and work design decisions, that is by choosing effective explainability approaches (XAI) to increase system transparency and interpretability for different stakeholders (see for example [19, 32]) and by allocation of functions between human and technology in ways that ensure complementarity between these two fundamentally different agents [11, 12, 44]. Existing tools from the HCI literature often focus solely on the visual design of interfaces or explanations for one target group (i.e., end users) [8, 40]. Even recently published guides to choosing explanations based on target groups, purpose of explanation, and AI life cycle phase fail to reflect the real-life complexity of high-risk use cases [42]. With ALIGN, control-accountability negotiations in complex use cases can be fostered and initial design ideas derived, which can then serve as a basis for further and more detailed decision making within AI projects.

### 5.1 Limitations, future research directions, and practical implications

While control-accountability alignment is central to all high-risk AI contexts, further research is needed to validate ALIGN in additional safety-critical domains, such as autonomous driving or healthcare, to challenge its applicability across all high-risk AI contexts. Nonetheless, the promising results of this initial evaluation across two safety-critical domains suggest that the method may be applicable to other high-risk AI contexts as well.

Also, our evaluation focused on the feasibility of the workshop concept and its perceived usefulness rather than on long-term organizational impact. As the timeline for AI use cases in safety-critical domain spans many years from exploration to system use, and most organizations in safety-critical domains currently do not yet have mature prototypes of AI systems but are still exploring the options for AI use in different contexts, it was not feasible to assess the effectiveness of ALIGN in terms of changed behaviors by stakeholders and improved outcomes of AI deployment. However, methods such as ALIGN may increase awareness and support stakeholders in proactively addressing questions of control and accountability, thereby fostering AI readiness and stakeholders' capabilities to effectively plan and manage AI use cases. The creation of a white paper as well as a research grant and proposal of a new research project by the regulatory agency for nuclear energy and the re-invitation for a second workshop underline the usefulness of ALIGN in fostering considerations of accountability and control in safety-critical domains. Furthermore the authors were informed several months after the workshop that, based on the scenarios used within the first workshop, the employees of the regulator had identified 7 relevant scenarios for their own application context that currently are being used for further internal discussions and as a base for the new research project.

Once there are more AI use cases in safety-critical domains, studies should examine how insights from the ALIGN workshop translate into actual design decisions, governance mechanisms, and safety outcomes. Both in the railways and in other domains, it will be crucial to show that supporting AI projects with ALIGN will result in better analysis and implementation of technical and work process requirements and finally in more effective and safe use of AI. Such research should also consider that the allocation of control and accountability may change as AI systems move from development to licensing, implementation and use. Therefore, the workshop should be conducted regularly within use cases and its effects evaluated longitudinally.

Our work has several practical implications for the design, implementation, use, and overall governance of high-risk AI systems. The ALIGN workshop serves as a tool to expand the focus on human-AI collaboration from end user interaction with a system and explainability to a holistic approach to systems design which draws on a range of concepts and methods to combine technology and work design. The scenario cards and canvas provide a way to make abstract concepts such as control and accountability accessible to heterogeneous stakeholders, including project leaders, executive managers, technical developers, human factors experts, regulators, and end users. Thereby, collaboration across disciplinary and organizational boundaries can be enhanced and the likelihood of sustainable alignment increased. ALIGN not only helps to identify where control-accountability gaps exist but also supports the creation of ideas for bridging them. For example, participants may identify the need for new explainability features, different function allocation between human and AI, or changes in organizational workflows to ensure that actors have the control necessary to meet their accountabilities. In this way, the method directly informs both technology and work design decisions.

By pushing organizations to document steps taken for optimized control-accountability alignment and the rationale behind design decisions, ALIGN also supports adherence to regulatory frameworks. The structured outputs of the workshop (i.e., design ideas for specific control-accountability misfits) can serve as documentation for internal audits or external certification processes. Thus, increasing the transparency of design decisions can help to convince AI regulators who act as gatekeepers for implementing new technologies in safety-critical domains. Regulatory requirements, such as those stated in the EU AI Act [45], often remain vague as to how they can be met through systems design. ALIGN provides practitioners with a concrete, structured way to identify and address control-accountability gaps early on in the design process and thereby answer to demands for enabling human oversight. Instead of addressing misalignments reactively,

ALIGN supports proactive work and technology design decisions that help to distribute control and accountability effectively across stakeholders. This is especially valuable in high-risk contexts, where late-stage corrections are particularly costly and associated with safety risks.

## 5.2 CONCLUSION

We have presented the new participatory design method ALIGN, which was inspired by both conceptual work on AI governance [11, 35] and empirical research within a real-life AI use case, and was evaluated through workshops with different stakeholders from safety-critical domains. ALIGN introduces scenario cards and an AI stakeholder canvas as ways to guide participants through assessing the control-accountability alignment in different safety-critical situations and generating design ideas for more effective handling of those situations. We thereby answer the call for tools to support the development of AI-based systems which enable effective human oversight in line with accountability assignments for different stakeholders. From a socio-technical perspective, ALIGN operates as a boundary-spanning intervention that brings together heterogeneous actors and translates between different forms of expertise, values, and constraints [1, 22]. Through scenario-based discussion, role reflection, and the joint development of design ideas, ALIGN makes implicit assumptions about control and accountability explicit and contestable. This process supports what Grote et al. [11] describe as decentralized and negotiated governance, in which alignment is not imposed top-down but emerges through integrative dialogue and perspective-taking. In practice, collaboration between different internal and external stakeholders within high-risk AI use cases is complex, requiring methods such as ALIGN to ensure equitable negotiations aimed at enabling stakeholders through control and motivating them through accountability to effectively and safely develop and use AI.

## ACKNOWLEDGMENTS

This research was funded by the ETH Zurich Foundation through the Future Mobility Research Program.